\documentclass[aps,prb,twocolumn,superscriptaddress,showpacs]{revtex4-1}
\usepackage{mathbbol,amsmath,amsfonts,amssymb,bm,color,dsfont}
\usepackage{graphicx,psfrag,array}

\usepackage{amsmath}
\usepackage{mathtools}
\usepackage{braket}
\usepackage{graphicx,psfrag,color}
\usepackage{dcolumn}
\usepackage{bm}
\usepackage{mathbbol, appendix}
\usepackage{multirow}
\usepackage{color,soul}
\usepackage{xcolor}
\DeclareMathOperator{\Tr}{Tr}

\providecommand{\ignore}[1]{}
\providecommand{\aucmnt}[1]{#1}
\def\be{\begin{equation}}
	\def\ee{\end{equation}}
\renewcommand{\aucmnt}[1]{}

\newcommand{\Comment}[1]{}

\newcommand{\Eq}[1]{Eq.~(\ref{#1})}

\begin{document}
\title{Bounds for low-energy spectral properties of center-of-mass conserving positive two-body interactions}
	
	\author{Amila Weerasinghe}
	\affiliation{Department of Physics, Washington University, St. Louis, Missouri 63130, USA}
	\author{Tahereh Mazaheri}
	\affiliation{Department of Physics, Washington University, St. Louis, Missouri 63130, USA}
	\author{Alexander Seidel}
	\affiliation{Department of Physics, Washington University, St. Louis, Missouri 63130, USA}
	\affiliation{Max-Planck-Institut f\"ur Physik Komplexer Systeme, N\"othnitzer Str. 38, 01187 Dresden, Germany}

	\date{\today}
	\begin{abstract}
		We study the low-energy spectral properties of positive center-of-mass conserving two-body Hamiltonians  as they arise in models of fractional quantum Hall states. 
		Starting from the observation that positive many-body Hamiltonians must have ground-state energies that increase monotonously in particle number, we explore what general additional constraints can be obtained for two-body interactions with ``center-of-mass conservation'' symmetry,
		both in the presence and absence of particle-hole symmetry. We find general bounds that constrain 
		the evolution of the ground-state energy with particle number, and in particular, constrain the chemical potential at $T=0$. Special attention is given to 
		Hamiltonians with zero modes, in which case similar bounds on the first excited state are also
		obtained, using a duality property. In this case, in particular, an upper bound on the charge gap is also obtained. We further comment on center of mass and relative decomposition in disk geometry
		within the framework of second quantization. 
	\end{abstract}
	\maketitle
	
	\section{Introduction}
	
	A cornerstone of the theory of fractional quantum Hall liquids is the construction and study of special parent Hamiltonians that stabilize prototypical wave functions such as the Laughlin state. The properties of such Hamiltonians have been well characterized analytically where their rich structure of so-called ``zero modes'' is concerned, i.e., states at zero, or the lowest possible, energy. These states are of fundamental importance to the physics of a quantum Hall phase, since, in known examples, they fully describe in particular the low-energy edge physics. In contrast, very little is known rigorously about the finite-energy properties of such special Hamiltonians and their more generic deformations.
	This article reports an effort at improving this situation.
	Our starting point is a general monotony property, in particle number, of the ground-state energy of positive many-body Hamiltonians. We observe that the strategy leading to this result 
	gives rise to
	further interesting bounds when combined with other properties of general interest in fractional quantum Hall model Hamiltonians, most importantly, center-of-mass conservation and the focus on two-body interaction. Our main result is a general bound on the 
	step size of the ground-state energy ( and in some cases the first excitation energy) with particle number. In its simplest version, it is obtained in situations with particle-hole symmetry but is subsequently improved and generalized to situations without particle-hole symmetry, including bosons. As a special application, an upper bound on the charge gap in special model Hamiltonians with zero modes is obtained. In the latter case, we also manage to give bounds on the evolution with particle number of the first excited state by observing a certain invariance property of the zero mode subspace and then introducing a dual version thereof.

	Technically, we work with second-quantized forms of projection-operator-type interactions.
	This is worth noting, since in this field, there is much history of deriving analytic results in a first quantized picture employing analytic wave functions\cite{laughlin,haldane_hierarchy,Halperin} and correspondingly constructed first quantized parent Hamiltonians.\cite{haldane_hierarchy, TK} As far as wave functions are concerned, their spectral decomposition in a particle number basis has become of interest in recent years through the study of the Jack-polynomial structure of special wave functions\cite{bernevig1, bernevig2} and through the more recently discovered matrix-product structure of these states.\cite{dubail, zelattel, bernevig_MP}
	In contrast, the use of second-quantized {\em Hamiltonians}, with some exceptions,\cite{LL}
	has long been reserved for numerical work, though their popularity  has recently increased as well, in part due to interest in fractional Chern insulators,\cite{FCI_sheng2011,qi, FCI_Chris,FCI_bernevig,FCI_Wen, scaffidi}
	purely technical reasons,\footnote{Such as setting up perturbative schemes, which is possible in the thin cylinder limit\cite{weerasinghe}}$^,$\cite{weerasinghe}
	as well as more general ones.\cite{ortiz, Li_paper, TMK_paper, Lee15}
	The preference for first quantized descriptions of parent Hamiltonians can perhaps be attributed to the fact that these are, by construction,  most suitable for studying the zero mode space, though it was recently shown (in some cases) that this is also possible in a purely second-quantized framework.\cite{ortiz, Li_paper, TMK_paper} Arguably, however, the advantage of working with first quantized Hamiltonians is lost when the focus is on finite-energy spectral properties.
	There, and moreover, when studying more generic Hamiltonians without any particularly interesting zero mode structure, arguments in favor of the greater efficiency of a ``pure guiding center'' description\cite{haldane11} are, in our opinion, particularly appealing. The second-quantized 
	presentation of Hamiltonians is one possible way to achieve such a pure guiding center description.  
	Our study can thus also be viewed as adding further emphasis to the utility of such an approach.

	\section{Monotony of Ground State Energy} \label{sec:Monotony}
	
	We begin by discussing the monotony in particle number and related general properties of the ground-state energy of positive many-body Hamiltonians. 
	To attain the desired level of generality, we will first consider a second-quantized $k$-body interaction
	of the form
	\be\label{k-body-V Hamilt}
	H_k =  \sum_{\substack{n_1,...,n_{2k}}}\,V_{n_1...n_{2k}}\,c_{n_1}^\dagger ... c_{n_k}^\dagger c_{n_{k+1}} ... c_{n_{2k}}\,.
	\ee
	The operators $c_n$ may satisfy bosonic or fermionic commutation relations.
	We will later focus on the special case where a ``center-of-mass'' conservation law
	is explicit, as is appropriate for model Hamiltonians of fractional quantum Hall
	type systems in various geometries. For the moment, however, the only additional property we will
	require is {\em positivity}, i.e., $\braket{\psi|H|\psi}\geq 0$ for all $k$-particle (and hence $N$-particle) kets $\ket{\psi}$.
	
	We now consider an $N$-particle mixed state described by a density matrix $\rho_N$.
	From $\rho_N$, we may define various $N'$-particle reduced density matrices $\rho_{N'}$,
	$N'<N$, given recursively via
	\be\label{reduced}
	\rho_{N'-1}=\frac 1 {N'} \sum_n c_n \rho_{N'} c_n^\dagger\,. 
	\ee
	We note that $\hat N\rho_{N'}=\rho_{N'}\hat N=N'\rho_{N'}$, where $\hat N=\sum_n c^\dagger_n c_n$
	is the particle number operator, and $\Tr \rho_{N'-1}=\Tr \rho_{N'}=1$.
	For both fermions and bosons, one easily verifies the relation
	\be\label{NH}
	\hat N H_k =k H_k + \sum_n c^\dagger_n H_k c_n\,,
	\ee
	obtained by commuting $c_n$ to the right.
	This gives
	\begin{align}
		\Tr \rho_{N'} H_k &= \frac{1}{N'} \Tr   \rho_{N'} \hat{N} H_k \nonumber\\
		&= \frac{k}{N'} \Tr   \rho_{N'} H_k + \Tr \rho_{N'-1} H_k \,,
	\end{align}
	or
	\be\label{induction}
	\Tr \rho_{N'-1} H_k = \frac{N'-k}{N'} \Tr \rho_{N'} H_k\,,
	\ee
	and by induction:
	\be
	\Tr \rho_{N'} H_k= \frac{(N-k)(N-1-k)\dotso(N'+1-k)}{N(N-1)\dotso (N'+1)}  \Tr \rho_{N} H_k.
	\ee
	We now denote the ground-state energy of $H_k$ in the $N$-particle sector
	as $E_0^k(N)$. Then choosing $\rho_N$ such that $\Tr \rho_{N} H_k=E_0^k(N)$,
	and noting $\Tr \rho_{N'} H_k\geq E_0^k(N')$ by the variational principle, we have
	\be\label{E0k}
	E^k_0(N')\leq \frac{(N-k)(N-1-k)\dotso(N'+1-k)}{N(N-1)\dotso (N'+1)}  E^k_0(N)\,.
	\ee
	
	So far we have not used positivity yet. 
	A result similar to \Eq{E0k} can also be obtained for general Hamiltonians 
	of the form
	\be\label{H}
	H=\sum_{k={k_{\sf min}}}^{k_{\sf max}} H_{k}\,,
	\ee
	where each term represents a positive ${k}$-body interaction, with $k_{\sf min}$ $(k_{\sf max}$) being the minimum (maximum)
	$k$.
	In this case we still have \Eq{induction} for each $H_{k}$, which, for positive interaction,
	in particular implies
	\be
	\Tr \rho_{N'-1} H_{k} \leq \frac{N'-k_{\sf min}}{N'}\, \Tr \rho_{N'} H_k\,,
	\ee
	and thus we have the same relation for $H$ in place of $H_k$.
	For the ground-state energy within the $N$-particle sector $E_0(N)$,
	we thus obtain \Eq{E0k} with $k_{\sf min}$ in place of $k$:
	\be
	\begin{split}
		\label{E0}
		& E_0(N')\leq   \\
		& \frac{(N-k_{\sf min})(N-1-k_{\sf min})\dotso(N'+1-k_{\sf min}) } {N(N-1)\dotso (N'+1)}
		E_0(N)\,.
	\end{split}
	\ee
	Clearly, this then implies in particular the monotony of the ground-state energy with particle number,
	\be\label{mono}
	E_0(N-1)\leq E_0(N)\,,
	\ee
	with equality {\em only} for $E_0(N)=0$.
	This result and a wealth of similar results all flow  from \Eq{E0} and have no doubt
	appeared previously in the literature, though we are unable to determine
	original references. For example, as another special case of \Eq{E0}, one obtaines the
	{\em superadditivity}\cite{lieb} of the ground-state energy. For this consider
	$N'=N_1$ and $N'=N_2$ with $N_1+N_2=N$, and add the corresponding instances
	of \Eq{E0}:
	\be
	E_0(N_1)+E_0(N_2) \leq ([N,N_1,k_{\sf min}]+[N,N_2,k_{\sf min}]) E_0(N)\,, 
	\ee
	where we have denoted the numerical factor in \Eq{E0} as
	$[N,N',k]=(N-k)!N'!/(N!(N'-k)!)$. It is easy to see that
	$[N,N_1,k_{\sf min}]+[N,N_2,k_{\sf min}]\leq 1$. To see this,
	one first observes that the left-hand side is equal to 1 for $k_{\sf min}=1$.
	Furthermore, $[N,N',k]$ monotonously decreases with increasing $k$.
	Hence we have the superadditivity
	\be
	E_0(N_1)+E_0(N_2) \leq E_0(N) \,,\; N_1+N_2=N\,.
	\ee
	
	At the level of generality assumed thus far, \Eq{E0} appears to be
	the strongest statement that can be made,  containing a
	multitude of 
	ground-state monotony properties 
	as special cases.
	In the following, we will be  interested in a more restricted but physically relevant 
	class of Hamiltonians
	that arises in particular when models of states in the fractional quantum Hall
	regime are considered. These Hamiltonians quite generally have an additional
	symmetry that in the second-quantized form \Eq{k-body-V Hamilt} manifests itself as ``center-of-mass'' conservation.\cite{seidel05} It turns out that in this case, further bounds 
	on the evolution of the ground-state energy with particle number can be given, and in some cases this is also true of the  first excited-state energy.
	
	\section{Specialization to center-of-mass conserving Hamiltonians}

	Many known parent Hamiltonians for various types of interesting fractional
	quantum Hall states have a peculiar way of satisfying \Eq{mono}: the ground-state energy $E_0(N)$ is exactly zero until the particle number reaches some value $N=N_I$,
	where $N_I/L$ approaches the incompressible filing factor and $L$ is the number of Landau level
	orbitals available to the system due to finite size geometry (e.g. finite disk, sphere, or torus).
	We will comment on the situation in the infinite disk geometry in Sec. \ref{infdisk}, where strictly
	speaking, absent any other constraints, $E_0(N)=0$ for any finite $N$ in the case of
	such special model Hamiltonians. It turns out that for quantum Hall type interaction
	Hamiltonians, additional constraints beyond \Eq{E0} can be given. 
	This is chiefly due to the general presence of another symmetry, that of
	the conservation of the center of mass. Related to that and in addition,
	some models of fermions possess a particle-hole symmetry.
	It is then natural to surmise that in cases where \Eq{mono} is saturated
	for $N<N_I$, another inequality should be saturated in the particle-hole
	symmetric region $N>L-N_I$. This turns out to be an {\em upper}
	bound on the step size in ground-state energy with particle number, $E_0(N)-E_0(N-1)$.
	This in particular provides an upper bound on the charge gap at the incompressible
	filling factor of special model Hamiltonians satisfying the ``zero mode paradigm''
	\be\label{zero}
	E_0(N)=0\;\;\text{for}\;\;N\leq N_I\,
	\ee
	but can be applied equally well to some more generic Hamiltonians.
	Before we derive these and other results, we will first write the Hamiltonian
	in a form in which all its pertinent properties are manifest. 
	
	In a constant magnetic background field, Landau-level projection leads to
	a one-dimensional ``lattice'' Hilbert space of single-particle orbitals labeled by
	an integer guiding center quantum number $n$, whose precise meaning 
	depends on the geometry and choice of basis. Here, these orbitals are
	created by the operators $c_n^\dagger$.
	Certain rotational and or (magnetic) translational symmetries manifest themselves
	as ``center-of-mass conservation'', i.e., the matrix element in \Eq{k-body-V Hamilt}
	is non zero only when $n_1+\dotsc n_k= n_{k+1}+\dotsc+n_{2k}$ is satisfied (on the torus 
	modulo $L$). This can be made manifest by writing the Hamiltonian \eqref{H}
	in the form
	\begin{subequations}\label{Hgen}
		\be
		H= \sum_{m=1}^M \sum_{r=0}^{k_m-1} \sum_{\substack{R\in \mathbb{Z}+r/k_m }} {Q_R^m}^\dagger Q_R^m\,,
		\ee
		where
		\be\label{QR}
		Q^m_R =\sum_{\substack{n_1,\dotso, n_{k_m}}} \!\!\!\eta^m_{R;n_1,\dotso,n_{k_m}} c_{n_1}\dotsm c_{n_{k_m}}\,.
		\ee
	\end{subequations}
	\Eq{Hgen} can be obtained from Eqs. \eqref{H} and \eqref{k-body-V Hamilt}
	by performing a spectral decomposition of the symbol $V_{n_1\dotsc n_{2k}}$,
	viewed as a big matrix with multi-indices $(n_1,\dotsc,n_k)$ and 
	$(n_{k+1},\dotsc,n_{2k})$. This matrix is block-diagonal in multi-indices of given
	$R=(n_1+\dotsc+n_k)/k$, and so eigenvectors $\eta^m_{R;n_1\dotsc n_k}$ can be labeled
	by $R$. These eigenvectors are normalized such that $\sum_{\{n_i\}} |\eta^m_{R;n_1\dotsc n_k}|^2$
	equals the corresponding eigenvalue, and the absence of negative coefficients signifies the positivity
	of the Hamiltonian. In \Eq{Hgen}, $M$ different terms labeled by $m$ are considered, each of which corresponds to an eigenvector of the aforementioned kind, obtained for a $k_m$-body operator in \Eq{H}, with non zero eigenvalue.
	To establish full equivalence between Eqs. \eqref{H} and \eqref{Hgen}, the case $M=\infty$ must be considered, whereas often $M$ will be finite in quantum Hall model Hamiltonians.
	In the following, we will refer to the Hamiltonian either in the form \eqref{Hgen}
	or in the less explicit but more condensed form Eqs. \eqref{k-body-V Hamilt}, \eqref{H},
	whichever is more convenient. We note that when working on the torus, 
	center-of-mass conservation strictly holds only ``modulo $L$''. In this case we 
	will still take the Hamiltonian to be of the form \Eq{Hgen}, where $c_n\equiv c_{n+L}$,
	and all symbols  $\eta^m_{R;n_1\dotsc n_k}$ are likewise invariant under the shift
	$n_i\rightarrow n_i+L$. 

	\subsection{Particle-hole symmetry}\label{gap}
	
	We will now demonstrate that the results already established
	have further  powerful implications on the evolution of the
	lowest eigenvalue with particle number
	in the presence of center-of-mass conservation as discussed above.
	A strikingly simple special instance of this is the case of $k=2$-body interactions
	for fermions. In this case, the spatial symmetries of the problem often
	also imply a particle-hole symmetry, as we will now discuss.

	We introduce the charge conjugation operator as a linear unitary operator
	$C$ defined via $Cc_nC=c_n^\dagger$, where $C=C^\dagger$, or $C^2=1\!\!1$.
	Consider now a two-body Hamiltonian $H_2$ as given in \Eq{k-body-V Hamilt},
	and assuming that the interaction matrix element $V_{n_1,n_2,n_3,n_4}$
	is proportional to $\delta_{n_1+n_2, n_3+n_4}$, easy calculation gives 
	\begin{subequations}\label{CHC}
		\be 
		CH_2C= 2  \sum_n \Delta_n -4 \sum_n \Delta_n c^\dagger_nc_n +H_2 \,,
		\ee  
		where
		\be
		\Delta_n= \sum_m  V_{mnnm}
		\ee
	\end{subequations}
	We emphasize that even though there is no strict center-of-mass conservation on the torus, but only
	``modulo $L$'',
	\Eq{CHC} is also obtained on the torus where the operators and $\eta$--form-factors have the 
	aforementioned periodicity. Specifically, if we write a {\em translationally invariant}
	two-body interaction on the torus in the form \eqref{Hgen} via
	\be\label{2body}
	Q^m_R=\frac 12 \sum_x \eta^m(x) c_{R-x} c_{R+x}\,, 
	\ee
	where $x$ runs over integer (half-odd-integer) values in the interval $[0,L)$ 
	for integer (half-odd-integer) $R$, and $\eta^m(x)$ satisfies $\eta^m(x+L)=\eta^m(x)$ along with,
	for fermions, $\eta^m(x)=-\eta^m(-x)$, we find
	\be\label{Delta}
	\Delta_n= \frac 1 4 \sum_m \sum_{x\in \frac 12 \mathbf{Z}} |\eta^m(x)|^2 \equiv \Delta,
	\ee
	where the sum is over all integer {\em and} half-odd-integer values in $[0,L)$.
	In particular, it is apparent that $\Delta_n\equiv \Delta$ does not depend on $n$ at all.
	The same result can also be obtained in the presence of spherical symmetry,
	where again it can be shown that\cite{Moeller}
	\be\label{Delta2}
	\Delta_n\equiv \Delta= \frac{1}{L}\sum_{m,n} V_{mnnm}\,.
	\ee
	We thus write \Eq{CHC} in its final form,
	\be \label{CHC2}
	CH_2C= 2 \Delta  L  -4 \Delta \hat N  + H_2 \,,
	\ee  
	which directly relates the spectrum at particle number $N$ to that at $L-N$.
	We read off:
	\be\label{ph}
	E_0(N)=E_0(L-N)+(4N-2L)\Delta\,.
	\ee
	\Eq{ph} applies not only to the ground-state energy but to the entire spectra at $N$ and $L-N$, respectively,
	and is the manifestation of particle-hole symmetry of fermionic two-body interactions 
	on the sphere or torus. It is straightforward to combine this last equation with the monotony result \Eq{mono} into a new bound on the step size of energy with particle number:
	\be\label{stepsize}
	E_0(N+1)-E_0(N)\leq 4\Delta\,.
	\ee
	In the thermodynamic limit, this in particular constrains the chemical potential at zero temperature.
	We emphasize, however, that the validity of \Eq{stepsize} is not limited to large system size. Also,
	we will in the following derive similar relations that can be applied to excitations. Hence we will refer to the quantity on the left-hand side of this equation by the more generic term ``energy step size'' in the following.
	
	Some remarks are in order to demonstrate that the above result is meaningful. 
	We may, for example, consider the $V_1$ Haldane pseudo-potential
	on the torus, with coefficients normalized such that
	\be
	\frac 1 2 \sum_{x\in \mathbf{Z} \;\mbox{or}\; x \in \mathbf{Z}+\frac 12} |\eta^1(x)|^2 \doteq 1\,,
	\ee
as befits a projection operator. (note double counting due to the fact that $x$ and $-x$ lead to identical terms in \Eq{2body}.) The $\doteq$ symbol signifies that on the torus, small deviations from the value of $1$ appear due to the standard periodization of pseudo-potentials, which vanish in the thermodynamic limit and are not present on the sphere. In either case, in the thermodynamic limit,
	\Eq{Delta} gives $\Delta=1$, owing to the fact that now $x$ runs over both integer and half-odd-integer values. In particular, the right-hand side of \Eq{stepsize} is of order unity.

Moreover, we observe that the inequality \Eq{stepsize} may be saturated.
If we sum over the first  $M$ odd Haldane pseudo-potentials, it is well known\cite{haldane_hierarchy}
that the resulting Hamiltonian satisfies \Eq{zero} with $N_I$ approaching $L/(M+1)$.
From \Eq{ph}, it is then clear that in this case,
\be\label{linearN}
    E_0(N)= (4N-2L)\Delta\;,\;\;\mbox{for $N\geq L-N_I$}\,.
\ee
Therefore, \Eq{stepsize} is the best possible bound on the energy step size
that is uniform in $N$. Below we will see that a slight improvement is possible
at the expense of bringing in more complicated, $N$-dependent coefficients.
The main benefit of the following considerations is, however, their greater generality.
We finally remark that 
the single-particle charge gap may be defined as $ \Delta_c = E_0(N+1)+E_0(N-1)-2E_0(N) $ . For the special (Laughlin state) parent Hamiltonians discussed above,
 the energy then jumps from $E(N\leq N_I)=0$ to the charge gap $E(N_I+1)\equiv \Delta_c$  at the incompressible
filling factor,
and we have in particular obtained an upper bound on this charge gap:
\be \label{Deltac}
    \Delta_c\leq 4\Delta\,.
\ee 

We note that this last relation is, in its functional form, reminiscent of results obtained
via the ``single mode approximation''. \cite{GMP1, GMP2} However, it essentially complements
the latter, which provides a variational upper bound on the {\em neutral} gap. 
It is further worth pointing out that the above was obtained solely by appealing to the
two principles of ground-state monotony and particle-hole symmetry, bypassing the need for
the construction of clever variational wave functions.

\subsection{Bosons, excited states, and duality }

The bound \Eq{stepsize} has the advantage of simplicity.  However, some limitations
thus far apply. So far, we have only considered particle-hole symmetric two-body interactions of fermions.
Interestingly, a road to generalization of this result manifests itself if we first limit our
attention to special model Hamiltonians and inquire about the evolution, in $N$, of the first excited state for such $N$ where zero modes are present (and the ground-state energy thus vanishes exactly). It turns out that this question can be investigated with methods similar to those of Sec. \ref{sec:Monotony}, thanks to the following fortuitous circumstance. It has been pointed out that
the zero mode space ${\cal H}_Z$ of the Hamiltonian is generally invariant under the action of
destruction operators $c_n$ (see, e.g., Refs. \onlinecite{Li_paper, TMK_paper}). Indeed, this follows
from the zero mode condition $Q^m_R\ket{\psi}=0$ for all $n$, $R$, and from the commutation relation $[Q_R^m,c_n]=0$. 
Much less appreciated seems to be the fact that there is a dual version of this statement.
Let ${\cal H}={\cal H}_Z\oplus {\cal H}_{NZ}$  be the decomposition of the Hilbert space
into the zero mode subspace and its orthogonal complement, the latter being spanned by all finite-energy eigenstates.
Then in fact ${\cal H}_{NZ}$ is invariant under the action of all creation operators $c_n^\dagger$.
For, if $\ket{\psi}\in {\cal H}_{NZ}$, and $\ket{\phi}$ is any zero mode, then $c_n\ket{\phi}$ is also a zero mode. Thus $\braket{\psi| c_n|\phi}=0=\braket{\phi| c_n^\dagger|\psi}$. So $c_n^\dagger\ket{\psi}$ is orthogonal to any zero mode. Thus $c_n^\dagger\ket{\psi}\in {\cal H}_{NZ}$ if $\ket{\psi}$ is.

We now consider $E_1(N,L)$, the lowest {\em non zero} eigenenergy for given particle number $N$, where in the following, we will make both the $N$ and the $L$ dependence explicit.
Note that for such $N$ where there are no zero modes, $E_1(N,L)=E_0(N,L)$, and the following considerations equally apply to this situation. In general, $E_1(N,L)$ is the ground-state energy of ${\cal H}_{NZ}$ for fixed $N$, and the invariance of ${\cal H}_{NZ}$ under the creation operators $c_n^\dagger$ allows us to proceed in a manner that  parallels the considerations  of Sec. \ref{sec:Monotony}, except stepping up in particle number instead of stepping down.

To this end, we restrict ourselves for now to two-body interactions of fermions, which we simply denote by $ H_2\equiv H $ below. For such interactions, we note the identity
\be\label{f2identity}
   (L-\hat N) H=\sum_n c_n c_n^\dagger H= 2H -4 \Delta \hat N +\sum_n c_n H c_n^\dagger\,,
\ee
where again center-of-mass conservation and symmetries have been used to extract the term $\Delta$, \Eq{Delta2}. It is then straightforward to proceed along the lines of Eqs. \eqref{NH}-\eqref{E0},
where only the concept of a reduced density matrix \Eq{reduced} must be replaced by 
a dual counterpart of an ``enlarged'' (in particle number) density matrix,
\be\label{enlarged}
    \rho_{N'+1}=\frac 1 {L-N'} \sum_n c_n^\dagger \rho_{N'} c_n\,. 
\ee
This then leads to the relation
\be\label{E1f}
E_1(N+1,L) -  \frac{L-N-2}{L-N}E_1(N,L) \leq \frac{4N}{L-N}\Delta\,.
\ee
We emphasize once more that \Eq{E1f} describes both the first excited-state energy in the presence of zero modes, as well as, in the absence of the latter, the ground-state energy.
As far as this second application is concerned, it is quite similar to \Eq{stepsize}, as the coefficient on the left-hand side is close to unity for large $L-N$, and the bound on the left-hand side even represents an improvement over \Eq{stepsize}  for $N<L/2$. One may again check that 
\Eq{E1f} is saturated in the regime discussed in and around \Eq{linearN}.

\Eq{E1f} has been derived by taking the expectation value of \Eq{f2identity} 
in the ground state of ${\cal H}_{NZ}$. If instead we again consider $N=N_I$,
the largest $N$ for which $E_0(N,L)=0$, and take the expectation value of \Eq{f2identity}
for the corresponding zero mode ground state, we obtain the following upper bound on the
charge gap:
\be\label{Deltac2}
      \Delta_c \leq \frac{4N_I}{L-N_I} \Delta\,,
\ee
which is usually (for $N_I/L<2$) an improvement over
\Eq{Deltac}.

We emphasize that while in deriving \Eq{f2identity}, 
we used the same symmetries that lead to particle-hole symmetry for fermions,
particle-hole symmetry does itself not seem to play any essential role here. To make
this point, we now derive analogous results for two-body interactions of bosons. 
In this case, the analog of \Eq{f2identity} is given by
\be
  (L+\hat N) H =-2H -4\Delta \hat N   +\sum_n c_n H c_n^\dagger\,.
\ee 
This then leads in an analogous manner to
\be\label{E1b}
   E_1(N+1,L) -  \frac{L+N+2}{L+N}E_1(N,L) \leq \frac{4N}{L+N}\Delta\,.
\ee
This is again similar in spirit to the ``step size'' equation \eqref{stepsize}, and, in addition to generalizing the latter to bosons, has the same benefit as \Eq{E1f}, applying also to the first excited state in the presence of zero modes. We may now further generalize  \Eq{Deltac2} to bosons via
\be\label{Deltac3}
 \Delta_c\leq \frac{4N_I}{L+N_I}\Delta\,.
\ee

We note one more subtle difference between Eqs. \eqref{E1f} and \eqref{E1b}.
In \Eq{E1f}, the positive term $2E_1/(L-N)$ may always be dropped if desired, in order to bound
the step size more directly. This is not immediately possible in \Eq{E1b}, where a similar term appears with opposite sign. While this term appears innocent at first, it gets somewhat out of control when
$E_1$ approaches order $L$. \textit{A priori}, we do not know when that happens. However, it is easy enough to use \Eq{E1b} in order to bound $E_1(N,L)$ directly. 
To demonstrate this, let us focus on the region $N>N_I$. In this case, we prove from \Eq{E1b} by easy induction
that
\be\label{E1bound}
    E_1(N,L)\leq \frac{2N(N-1)}{L+1}\Delta \quad(N>N_I)\,,
\ee
using \Eq{Deltac3} with $N=N_I+1$ as the starting point of the induction.
It is clear that \Eq{E1bound} may be much improved if $N_I$ is appreciably larger than $1$.
Proceeding with the general \Eq{E1bound}, however, we in particular see that 
\be
 E_0(N,L)\leq 2L\Delta \quad(N<L)\,,
\ee
where we note again that $E_1$ and $E_0$ are defined to be the same for $N>N_I$.
This in \Eq{E1b} actually reproduces the original step size equation \eqref{stepsize}, now for bosons, with the additional
restriction of $N\leq L$, i.e., filling factor no greater than $1$.

In closing this section, we evaluate the bound \eqref{Deltac2}
for the important special case of the
$V_1$ Haldane pseudo-potential on a sphere threaded by $N_\Phi=L-1$
flux quanta, which stabilizes a $\nu=1/3$ Laughlin state of
 $N_I$ particles, where $N_\Phi=3(N_I-1)$. Table \ref{tab} summarizes our results.
Note that as defined above in \Eq{Deltac}, the charge gap $\Delta_c$ corresponds to a state of $N=N_I+1$ particles or the insertion of three quasi-particles into the Laughlin state, which must be well separated before the thermodynamic limit is reached. While this is not quite the case for the system sizes shown in Table
\ref{tab} yet, we emphasize that the bounds derived here apply equally well to finite particle number. Moreover, one may be confident from the data given that even in the thermodynamic limit, our upper bound \Eq{Deltac2} overestimates the charge gap by less than a factor of 2. This seems quite reasonable, given the great generality of the
bounds derived here. As we stressed above, these bounds are saturated in certain cases; hence there is not much room for improvement at this level of generality. 
In this light, the fact that \Eq{Deltac2} is within less than a factor of 2 of the actual gap
seems quite satisfactory.

\begin{table}
	\renewcommand{\arraystretch}{1.2}
	\begin{tabular}{m{1cm} m{1cm} m{1.5cm} m{1.5cm} m{1.5cm} m{1.2cm}}
		\hline
		\hline
		\textit{N}         & \textit{L}      & $ \Delta $ & $ (\Delta_c)_{ub} $ & $ (\Delta_c)_{ED} $ & $ \dfrac{(\Delta_c)_{ub}}{(\Delta_c)_{ED}} $ \\[0.4cm]
		\hline
		5         & 10     & 0.8500     & 2.2667         & 1.6406	& 1.38 \\
		6         & 13     & 0.8846     & 2.2115         & 1.4600	& 1.52 \\
		7         & 16     & 0.9063     & 2.1750         & 1.4208	& 1.53 \\
		8         & 29     & 0.9211     & 2.1491         & 1.3692	& 1.57 \\
		9         & 22     & 0.9318     & 2.1299         & 1.3367	& 1.60 \\
		10        & 25     & 0.9400     & 2.1150         & 1.3100	& 1.62 \\
		\hline
		\hline
	\end{tabular}
	\caption{\label{tab}Charge gap for a sphere threaded by $N_\Phi=L-1$ flux quanta, at the incompressible filling factor of the $V_1$ Haldane pseudo-potential (see text).
$N$ represents particle number, the parameter $\Delta$ defined by \Eq{Delta2}
equals $1- 3/(2L)$ for the $V_1$ pseudopotential, $(\Delta_c)_{ub}$ is the upper bound on the charge gap as given by \Eq{Deltac2}, and $(\Delta_c)_{ED}$ is the actual charge gap as determined by exact diagonalization.\label{table:gap}}
\end{table}
 
\subsection{Considerations for the disk}\label{infdisk}

Most of the above results, except for the general monotony \Eq{mono}, are not in any obvious way applicable or sensible in the infinite disk geometry, where $L=\infty$.
In this case, the Hilbert space of any rotationally invariant Hamiltonian nonetheless 
decomposes into finite dimensional subspaces of given particle number $N$ {\em and} 
given angular momentum ${\cal L}_z$. 
The lowest energy $E_0(N, {\cal L}_z)$ then satisfies a fairly obvious monotony
relation in the angular momentum variable ${\cal L}_z$, which we wish to mention here
for completeness.

In the disk geometry, a decomposition into center of mass and relative degrees of freedom is possible, and any Hamiltonian with translational and rotational invariance will decouple from
the center-of-mass degrees of freedom. In this context, it is useful to introduce ladder operators
$a_i$, $a_i^\dagger$, $[a_i, a_j^\dagger]=\delta_{ij}$ such that $a^\dagger_ia_i$ is the angular momentum of the $i$ th particle. We stress that these operators are very different from the
``second-quantized'' operators $c_n$, which carry orbital indices and preserve the symmetry of the
wave function. In contrast, the $a_i$ carry particle indices like any first quantized single-particle operators, thus not by themselves preserving the symmetry of the wave function, which is also
not in any way encoded in the commutation relations of the $a_i$. As a result, the following 
is independent of particle statistics.

In this description, the relative degrees of freedom are (over-)completely described by
the operators $a_i-a_j$ and their Hermitian adjoints. The total angular momentum operator
may be decomposed into a center-of-mass part and a relative part, respectively, via
\be \label{lsum}
     {\cal L}_z={\cal L}_z^{\sf C} + {\cal L}_z^{\sf rel} \,.
\ee 
The operator $b=\frac{1}{\sqrt{N}} \sum_i a_i$ and its adjoint $b^\dagger$ commute with all
$a_i-a_j$, $a_i^\dagger-a_j^\dagger$, and thus with the Hamiltonian and with ${\cal L}_z^{\sf rel}$.
 Clearly, also, $b^\dagger$ raises ${\cal L}_z=\sum_i a_i^\dagger a_i$, and thus ${\cal L}_z^C$, by $1$. $b^\dagger $ and $b$ are thus ladder operators for the center-of-mass part of the angular momentum. From the commutation relation
 \be\label{bcomm}
    [b,b^\dagger]=1
 \ee
 it follows that $b^\dagger$ cannot annihilate any non-zero ket of the Hilbert space.
 From the above it thus follows that the entire spectrum $\Sigma_{{\cal L}_z}$ for a given value of ${\cal L}_z$
 is contained in that for ${\cal L}_z+1$, $\Sigma_{{\cal L}_z+1}$: $b^\dagger$ always raises the value of ${\cal L}_z$
 while keeping the energy the same. This in particular implies the monotony
 \be\label{lmono}
    E_0(N,{\cal L}_z) \geq E_0(N,{\cal L}_z+1)\,.
 \ee 
The above is simply a manifestation of center-of-mass degeneracy in the infinite plane,
and is not too surprising. Note that unlike when using conjugate magnetic translations to establish
a similar degeneracy (cf., e.g., Ref. \onlinecite{seidel05}), the equality of the spectra at  ${\cal L}_z$ and ${\cal L}_z+1$ does not follow, since $b$ can, in general, annihilate non zero kets.  However, $\Sigma_{{\cal L}_z}$ is identical to the spectrum
associated to  the subspace having angular momentum ${\cal L}_z+1$ and ${\cal L}^C_z>0$.
In determining the full spectrum, it is thus sufficient to focus on
the subspaces characterized by ${\cal L}^C_{z}=0$ and all possible values for ${\cal L}_z={\cal L}_z^{\sf rel}$. 
A more interesting question is thus whether the ground-state energy $E_0(N, {\cal L}_z, {\cal L}_z^C=0)$ as a function of given $N$, ${\cal L}_z$ and subject to the constraint  ${\cal L}_z^C=0$ satisfies a monotony similar to \Eq{lmono}. 
We leave detailed analysis as an interesting problem for the future.

We emphasize that the above is true exactly only when no cutoff in orbital space is imposed, other than what naturally follows from fixing total angular momentum (i.e., for bosons, no orbitals with $n>{\cal L}_z$ allowed, and a correspondingly lower cutoff for fermions). 
Since, for example, in numerical calculations the second-quantized framework used throughout most of this paper may be deemed preferable, we will give second-quantized expressions for the operators $b$, $b^\dagger$ and the various components of  angular momentum appearing in \Eq{lsum}.

Clearly, the operators $b$, $b^\dagger$ are single body operators changing the total angular momentum by $\pm 1$ while preserving energy. In particular, they preserve zero modes of special
Hamiltonians. This last circumstance allows us to
make contact with our recent work,\cite{ortiz,  TMK_paper, Li_paper} where a class of second-quantized single body operators was discussed for various geometries that preserve zero modes.
These operators ${\cal O}_d$ are labeled by an integer $d$, and in disk geometry, raise ${\cal L}_z$ by $d$. Up to some arbitrary normalization which we will fix here for our purposes, the operator ${\cal O}_1$ for the disk is given by\cite{ortiz}
\be\label{O1}
  {\cal O}_1 = \sum_{n=0}^\infty \sqrt{n+1} \,c^\dagger_{n+1} c_n\,.
\ee
It is thus natural to assume that $b^\dagger$ is proportional to this operator. Indeed,
the action of ${\cal O}_1$ can be seen,\cite{ortiz, TMK_paper} at first quantized level, to correspond
to multiplication of an analytic wave function with a factor proportional to $\sum_{i=1}^N z_i$.
Here, $z_i=x_i+iy_i$ as usual. The same can easily be established for the operator $b^\dagger$.
Hence up to normalization, these operators are the same. One easily verifies
$[{\cal O}_1^\dagger, {\cal O}_1] = \hat N$, such that comparison with \Eq{bcomm} gives
\be
   b^\dagger = \frac{1}{\sqrt{\hat N}} {\cal O}_1\,.
\ee
Alternatively, one may verify directly, if desired, that \Eq{O1} commutes with all Haldane pseudo-potentials, and thus only acts on center-of-mass degrees of freedom. 
The operators for various aspects of angular momentum are then given by
\be
 {\cal L}_z=  \sum_{n=0}^\infty n \,c^\dagger_{n} c_n\,,
\ee
which is obvious. Furthermore,
\be
   {\cal L}_z^C = b^\dagger b\,,
\ee
which is, up to terms proportional to particle number, a two-body operator, and
${\cal L}^{\sf rel}_z$ is then obtained from \Eq{lsum}.

\section{Discussion and Conclusion}\label{discussion}

In this paper, we have been interested in bounds describing the evolution in particle number of low-energy spectral properties for both general and ``special'' positive two-body interaction Hamiltonians.
Even in the latter, ``special'' case, we have accessed properties which are beyond the zero mode subspace that renders these Hamiltonians special. We have thus naturally employed a second-quantized framework, which we find superior for addressing any properties not directly related to zero modes, whether or not the latter are present.

Our starting point has been the general monotony of the ground-state energy in particle number, for any positive interaction Hamiltonian. We then asked what additional information can be obtained 
using strategies related to those used in the proof of this monotony property when additional 
assumptions applying to a wide class of fractional quantum Hall model Hamiltonians are made. Specifically, we have focused on two-body interactions with center-of-mass conservation.
For fermions, in many situations of interest one also has a particle-hole symmetry, and we observed that this alone can be used to immediately translate the monotony property into a general bound on the energy step size, giving in particular an upper bound on the charge gap 
at the ``incompressible filling factor'' of special model Hamiltonians. This bound was subsequently improved and generalized to bosons. We also used a dual argument to obtain similar bounds for the first excited state in the presence of zero modes, thus showing that for special model Hamiltonians possessing the latter, non-trivial statements are possible even for excited states at non zero energy.

While we have been mostly concerned with compact geometries, in particular, the torus and sphere,
we have also commented on the situation in the infinite disk geometry, where center-of mass degeneracy leads to an obvious monotony property as a function of angular momentum. In this context, we also commented on aspects of center-of-mass and relative-coordinate decomposition in the framework of second quantization. 

We are hopeful that these results will spur further development regarding exact properties beyond zero modes in model Hamiltonians of fractional quantum Hall states and related systems.

\begin{acknowledgments}
This work has been supported by the National Science
Foundation under NSF Grant No. DMR-1206781. A.S. would like to thank the Max Plank Institut f\"ur 
Physik Komplexer Systeme for hospitality, where part of this work has been conducted. A.S. also thanks X. Tang, G. Ortiz, and Z. Nussinov for insightful discussion of related topics. 
\end{acknowledgments}

\bibliography{monotonybib_PRB}{}

\begin{thebibliography}{10}

\bibitem{laughlin}
R.~B. Laughlin.
\newblock {Anomalous Quantum Hall Effect: An Incompressible Quantum Fluid with
  Fractionally Charged Excitations}.
\newblock {\em Phys. Rev. Lett.}, 50:1395--1398, May 1983.

\bibitem{haldane_hierarchy}
F.~D.~M. Haldane.
\newblock {Fractional Quantization of the Hall Effect: A Hierarchy of
  Incompressible Quantum Fluid States}.
\newblock {\em Phys. Rev. Lett.}, 51:605--608, Aug 1983.

\bibitem{Halperin}
B.~I. Halperin.
\newblock {Statistics of Quasiparticles and the Hierarchy of Fractional
  Quantized Hall States}.
\newblock {\em Phys. Rev. Lett.}, 52:1583--1586, Apr 1984.

\bibitem{TK}
S.~A. Trugman and S.~Kivelson.
\newblock {Exact results for the fractional quantum Hall effect with general
  interactions}.
\newblock {\em Phys. Rev. B}, 31:5280--5284, Apr 1985.

\bibitem{bernevig1}
B.~Andrei Bernevig and F.~D.~M. Haldane.
\newblock {Model Fractional Quantum Hall States and Jack Polynomials}.
\newblock {\em Phys. Rev. Lett.}, 100:246802, Jun 2008.

\bibitem{bernevig2}
B.~Andrei Bernevig and F.~D.~M. Haldane.
\newblock {Generalized clustering conditions of Jack polynomials at negative
  Jack parameter $\ensuremath{\alpha}$}.
\newblock {\em Phys. Rev. B}, 77:184502, May 2008.

\bibitem{dubail}
J.~{Dubail}, N.~{Read}, and E.~H. {Rezayi}.
\newblock {Edge-state inner products and real-space entanglement spectrum of
  trial quantum Hall states}.
\newblock {\em Physical Review B}, 86(24):245310, December 2012.

\bibitem{zelattel}
Michael~P. Zaletel and Roger S.~K. Mong.
\newblock {Exact matrix product states for quantum Hall wave functions}.
\newblock {\em Phys. Rev. B}, 86:245305, Dec 2012.

\bibitem{bernevig_MP}
B.~Estienne, Z.~Papi\ifmmode~\acute{c}\else \'{c}\fi{}, N.~Regnault, and B.~A.
  Bernevig.
\newblock {Matrix product states for trial quantum Hall states}.
\newblock {\em Phys. Rev. B}, 87:161112, Apr 2013.

\bibitem{LL}
Dung-Hai Lee and Jon~Magne Leinaas.
\newblock {Mott Insulators without Symmetry Breaking}.
\newblock {\em Phys. Rev. Lett.}, 92:096401, Mar 2004.

\bibitem{FCI_sheng2011}
DN~Sheng, Zheng-Cheng Gu, Kai Sun, and L~Sheng.
\newblock {Fractional quantum Hall effect in the absence of Landau levels}.
\newblock {\em Nature communications}, 2:389, 2011.

\bibitem{qi}
Xiao-Liang Qi.
\newblock {Generic Wave-Function Description of Fractional Quantum Anomalous
  Hall States and Fractional Topological Insulators}.
\newblock {\em Phys. Rev. Lett.}, 107:126803, Sep 2011.

\bibitem{FCI_Chris}
Titus Neupert, Luiz Santos, Claudio Chamon, and Christopher Mudry.
\newblock {Fractional Quantum Hall States at Zero Magnetic Field}.
\newblock {\em Phys. Rev. Lett.}, 106:236804, Jun 2011.

\bibitem{FCI_bernevig}
N.~Regnault and B.~Andrei Bernevig.
\newblock {Fractional Chern Insulator}.
\newblock {\em Phys. Rev. X}, 1:021014, Dec 2011.

\bibitem{FCI_Wen}
Evelyn Tang, Jia-Wei Mei, and Xiao-Gang Wen.
\newblock {High-Temperature Fractional Quantum Hall States}.
\newblock {\em Phys. Rev. Lett.}, 106:236802, Jun 2011.

\bibitem{scaffidi}
T.~{Scaffidi} and S.~H. {Simon}.
\newblock {{Exact Solutions of Fractional Chern Insulators: Interacting
  Particles in the Hofstadter Model at Finite Size}}.
\newblock {\em Phys. Rev. B}, 90:115132, July 2014.

\bibitem{Note1}
Such as setting up perturbative schemes, which is possible in the thin cylinder
  limit\cite {weerasinghe}.

\bibitem{weerasinghe}
Amila Weerasinghe and Alexander Seidel.
\newblock {Thin torus perturbative analysis of elementary excitations in the
  Gaffnian and Haldane-Rezayi quantum Hall states}.
\newblock {\em Phys. Rev. B}, 90:125146, Sep 2014.

\bibitem{ortiz}
G.~Ortiz, Z.~Nussinov, J.~Dukelsky, and A.~Seidel.
\newblock {Repulsive interactions in quantum Hall systems as a pairing
  problem}.
\newblock {\em Phys. Rev. B}, 88:165303, Oct 2013.

\bibitem{Li_paper}
Li~Chen and Alexander Seidel.
\newblock {Algebraic approach to the study of zero modes of Haldane
  pseudopotentials}.
\newblock {\em Phys. Rev. B}, 91:085103, Feb 2015.

\bibitem{TMK_paper}
Tahereh Mazaheri, Gerardo Ortiz, Zohar Nussinov, and Alexander Seidel.
\newblock {Zero modes, bosonization, and topological quantum order: The
  Laughlin state in second quantization}.
\newblock {\em Phys. Rev. B}, 91:085115, Feb 2015.

\bibitem{Lee15}
Ching~Hua Lee, Zlatko Papi\ifmmode~\acute{c}\else \'{c}\fi{}, and Ronny
  Thomale.
\newblock {Geometric Construction of Quantum Hall Clustering Hamiltonians}.
\newblock {\em Phys. Rev. X}, 5:041003, Oct 2015.

\bibitem{haldane11}
F.~D.~M. Haldane.
\newblock {Geometrical Description of the Fractional Quantum Hall Effect}.
\newblock {\em Phys. Rev. Lett.}, 107:116801, Sep 2011.

\bibitem{lieb}
ElliottH. Lieb.
\newblock {The Bose Gas: A Subtle Many-Body Problem}.
\newblock In Walter Thirring, editor, {\em The Stability of Matter: From Atoms
  to Stars}, pages 699--719. Springer Berlin Heidelberg, 2001.

\bibitem{seidel05}
Alexander Seidel, Henry Fu, Dung-Hai Lee, Jon~Magne Leinaas, and Joel Moore.
\newblock {Incompressible Quantum Liquids and New Conservation Laws}.
\newblock {\em Phys. Rev. Lett.}, 95:266405, Dec 2005.

\bibitem{Moeller}
Gunnar M\"oller and Steven~H. Simon.
\newblock {Composite fermions in a negative effective magnetic field: A Monte
  Carlo study}.
\newblock {\em Phys. Rev. B}, 72:045344, Jul 2005.

\bibitem{GMP1}
S.~M. Girvin, A.~H. MacDonald, and P.~M. Platzman.
\newblock {Collective-Excitation Gap in the Fractional Quantum Hall Effect}.
\newblock {\em Phys. Rev. Lett.}, 54:581--583, Feb 1985.

\bibitem{GMP2}
S.~M. Girvin, A.~H. MacDonald, and P.~M. Platzman.
\newblock {Magneto-roton theory of collective excitations in the fractional
  quantum Hall effect}.
\newblock {\em Phys. Rev. B}, 33:2481--2494, Feb 1986.

\end{thebibliography}
\bibliographystyle{unsrt}


\end{document}